\documentclass[twocolumn,pre,superscriptaddress]{revtex4}

\usepackage{dcolumn}
\usepackage{amsmath}

\usepackage{graphicx}

\begin{document}

\renewcommand{\d}{\mathrm{d}}
\newcommand{\ii}{\mathrm{i}}
\newcommand{\Ord}{\mathrm{O}}
\newcommand{\e}{\mathrm{e}}
\newcommand{\half}{\mbox{$\frac12$}}
\newcommand{\set}[1]{\lbrace#1\rbrace}
\newcommand{\av}[1]{\langle#1\rangle}
\newcommand{\etal}{{\it{}et~al.}}
\newcommand{\defn}{\textit}
\newcommand{\stirling}[2]{\biggl\lbrace{\!{#1\atop#2}\!}\biggr\rbrace}

\newlength{\figurewidth}
\setlength{\figurewidth}{0.95\columnwidth}
\setlength{\parskip}{0pt}
\setlength{\tabcolsep}{6pt}
\setlength{\arraycolsep}{2pt}

\title{Random hypergraphs and their applications}
\author{Gourab Ghoshal}
\affiliation{Department of Physics and Michigan Center for Theoretical Physics, University of Michigan, Ann Arbor, Michigan 48109, USA}
\author{Vinko Zlati\'{c}\thanks{vzlatic@irb.hr}}
\affiliation{Theoretical Physics Division, Rudjer Bo\v{s}kovi\'{c} Institute, P.O.Box 180, HR-10002 Zagreb, Croatia}
\affiliation{CNR-INFM Centro SMC Dipartimento di Fisica, Universit\`a di Roma ``Sapienza'' P.le Moro 5, 00185 Roma, Italy }
\author{Guido Caldarelli\thanks{Guido.Caldarelli@roma1.infn.it}}
\affiliation{CNR-INFM Centro SMC Dipartimento di Fisica, Universit\`a di Roma ``Sapienza'' P.le Moro 5, 00185 Roma, Italy }
\affiliation{Linkalab, Complex Systems Computational Lab. 09100 Cagliari Italy}
\author{M. E. J. Newman}
\affiliation{Department of Physics and Center for the Study of Complex Systems, University of
Michigan, Ann Arbor, Michigan 48109, USA}
\affiliation{Santa Fe Institute, Santa Fe, New Mexico 87501, USA}
\begin{abstract}
  In the last few years we have witnessed the emergence, primarily in
  on-line communities, of new types of social networks that require for
  their representation more complex graph structures than have been
  employed in the past.  One example is the folksonomy, a tripartite
  structure of users, resources, and tags---labels collaboratively applied
  by the users to the resources in order to impart meaningful structure on
  an otherwise undifferentiated database.  Here we propose a mathematical
  model of such tripartite structures which represents them as random
  hypergraphs.  We show that it is possible to calculate many properties of
  this model exactly in the limit of large network size and we compare the
  results against observations of a real folksonomy, that of the on-line
  photography web site Flickr.  We show that in some cases the model
  matches the properties of the observed network well, while in others
  there are significant differences, which we find to be attributable to
  the practice of multiple tagging, i.e.,~the application by a single user
  of many tags to one resource, or one tag to many resources.
\end{abstract}

\pacs{89.75.Fb, 89.75.Hc}

\maketitle

\section{Introduction}
\label{sec:intro}
Networks are a versatile mathematical tool for representing the structure
of complex systems and have been the subject of large volume of work in the
last few
years~\cite{Newman_2003_01,BLMCH_2006,DM_book_2003,NBW_book_2006,Caldarelli07}.
In its simplest form a network consists of a set of nodes or vertices,
connected by lines or edges, but many extensions and generalizations have
also been studied, including networks with directed edges, networks with
labeled or weighted edges or vertices, and bipartite networks, which have
two types of vertices and edges running only between unlike types.

Recently, however, new and more complex types of network data have become
available, especially associated with on-line social and professional
communities, that cannot adequately be described by existing network
formats.  One example is the \emph{folksonomy}.  ``Folksonomy'' is the name
given to the common on-line (and sometimes off-line) process by which a
group of individuals collaboratively annotate a data set to create semantic
structure.  Typically mark-up is performed by labeling pieces of data with
\defn{tags}.  A good example is provided by the on-line photography
resource Flickr, a web site to which users upload photographs that can then
be viewed by other users.  Flickr allows any user to give a short
description of any photo they see, usually just a single word or a few
words.  These are the tags.  In principle, tags can allow users to do many
things, such as searching for photos with particular subjects or clustering
photos into topical groups.  There are also many other websites and on-line
resources with similar tagging capabilities, but dealing with different
resources.  On the website CiteUlike, for example, users upload academic
papers as opposed to photographs and label them with descriptive tags.

Researchers have taken a variety of approaches to the representation of
folksonomy data using network methods, including modeling them as simple
unipartite graphs and bipartite graphs as well as limited forms of
tripartite
graphs~\cite{PINTS,Sapienza_folkpaper,Ausloos_Lambiotte_2006,PFPDV_2008}.
Each of these approaches, however, fails to capture some elements of the
structure of the data and hence limits the conclusions that can be drawn
from subsequent network analysis.

The fundamental building block in a folksonomy is a triple consisting of a
\defn{resource}, such as a photograph, a \defn{tag}, usually a short text
phrase, and a \defn{user}, who applies the tag to the resource.  Any full
network representation of folksonomy data needs to capture this three-way
relationship between resource, tag, and user, and this leads us to the
consideration of hypergraphs.

\begin{figure}[t]
\includegraphics[width=5cm]{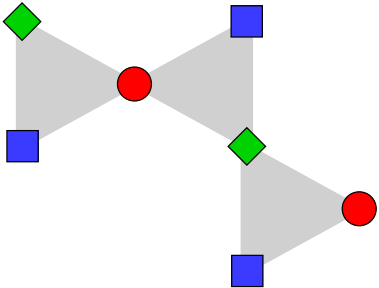}
\caption{Vertices in our networks come in three types, represented here by
  the red circles, green diamonds, and blue squares, and are connected by
  three-way hyperedges that each join together exactly one circle, one
  diamond, and one square.  In the language of folksonomies, the circles
  represent, say, the resources, the diamonds the tags, and the squares the
  users.}
\label{fig:hyperedges}
\end{figure}

A \defn{hypergraph} is a generalization of an ordinary graph in which an
edge (or \defn{hyperedge}) can connect more than two vertices together.  To
represent our folksonomy we make use of a \defn{tripartite hypergraph}, a
generalization of the more familiar bipartite graph, in which there are
three types of vertices representing resources, tags, and users, and
three-way hyperedges joining them in such a way that each hyperedge links
together exactly one resource, one tag, and one user.  Each hyperedge
corresponds to the act of a user applying a tag to a resource and hence the
tripartite hypergraph preserves the full structure of the folksonomy---see
Fig.~\ref{fig:hyperedges}.

In this paper, we study the theory of such tripartite graphs, starting with
basic network properties such as degree distributions and then developing a
random graph model that allows us to make analytic predictions of a variety
of network properties.  We test our predictions by comparing them with data
from the Flickr folksonomy and find good agreement in some, but not all,
cases.

\section{Tripartite graphs}
\label{sec:tripartite}
We begin our study of tripartite hypergraphs by outlining some of the basic
properties of such networks.  Our tripartite graphs have three different
types of vertices, which, to preserve generality, we will refer to as red,
green, and blue vertices.  (In this paper, when discussing applications of
the theory to folksonomies, red will represent resources, green tags, and
blue users, but the theory itself is entirely agnostic about what the
colors represent.)  Let us suppose that there are $n_r$ red vertices,
$n_g$~green ones, and $n_b$ blue ones.

The edges in our network are three-way hyperedges that each connect one
red, one green, and one blue vertex.  (We might say that the hyperedges are
``colorless'' or ``white,'' since red, green, and blue make white when
combined in the human visual system.)  Let us suppose there to be $m$
hyperedges in total.

There are a number of ways in which vertex degree can be defined for a
hypergraph.  Some authors, for instance, have defined degree as the total
number of other vertices to which a given vertex is connected by
hyperedges.  This corresponds to the definition of degree in an ordinary
graph (at least when there are no multiedges or self-edges), but in failing
to distinguish between the different types of vertices to which hyperedges
are connected, it can lead to confusion in the hypergraph case.  The best,
and also simplest, definition of degree for a vertex in a hypergraph is
simply the number of hyperedges attached to that vertex.  Thus a red vertex
participating in four hyperedges has degree four.  This might mean that it
has four green and four blue neighbors in the network, but it is also
possible that some neighboring vertices are common to more than one
hyperedge, in which case the number of neighboring vertices of a given
color may be smaller than four.

The mean degree~$c_r$ of a red vertex in our network is given by the number
of hyperedges in the network divided by the number of red vertices, and
similarly for green and blue:
\begin{equation}
c_r = {m\over n_r},\qquad
c_g = {m\over n_g},\qquad
c_b = {m\over n_b}.
\end{equation}
Rearranging these equations to give three separate expressions for~$m$, we
also have,
\begin{equation}
n_r c_r = n_g c_g = n_b c_b = m.
\label{eq:constraint}
\end{equation}
Thus the mean degrees of the different vertex types cannot be chosen
independently, but are linked via the fact that the same hyperedges connect
to the red, green and blue vertices.

One of the most important parameters of a network is its degree
distribution.  Just as bipartite networks have two distinct degree
distributions, our tripartite ones have three: we define $p_r(k)$ to be the
fraction of red vertices in the network that have degree~$k$, and $p_g(k)$
and~$p_b(k)$ to be the corresponding quantities for green and blue
vertices.  These distributions satisfy the sum rules
\begin{equation}
\sum_{k=0}^\infty p_r(k) = \sum_{k=0}^\infty p_g(k)
  = \sum_{k=0}^\infty p_b(k) = 1,
\label{eq:sumrule1}
\end{equation}
and
\begin{equation}
\sum_{k=0}^\infty k p_r(k) = c_r,\quad
\sum_{k=0}^\infty k p_g(k) = c_g,\quad
\sum_{k=0}^\infty k p_b(k) = c_b.
\label{eq:sumrule2}
\end{equation}

As with bipartite graphs, it is sometimes convenient to form
``projections'' of tripartite graphs onto a subset of their vertices.  In a
bipartite graph of red and green vertices, for instance, one forms a
projection onto the red vertices alone by constructing the network of red
vertices in which vertices are connected by an edge if they share a common
green neighbor in the original bipartite graph~\cite{NSW01}.

\begin{figure}[t]
\includegraphics[width=5.5cm]{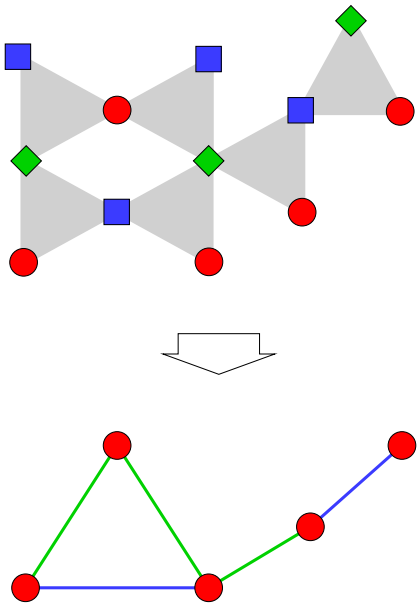}
\caption{Ways of projecting a tripartite graph onto one of its vertex types
  (red in this case).  Red vertices in the projected graph can be connected
  if they share a green neighbor (green edges in the projected graph), a
  blue neighbor (blue edges), or a neighbor of either kind (all edges
  together).}
\label{fig:project}
\end{figure}

While for bipartite graphs there is essentially only one way of performing
projections, there are several distinct possibilities for tripartite
graphs---see Fig.~\ref{fig:project}.  One can again join two red vertices
if they share a green neighbor---in our Flickr example from the
introduction, two photos would be connected if they have a tag in common.
Or one can join two red vertices that share a common blue neighbor---two
photos that were tagged by the same user.  Or one could join vertices that
share \emph{either} a green or a blue neighbor.  And of course one can
define the equivalent projections onto the green and blue vertices.

But it doesn't stop there.  In a tripartite network, one can also form
projections onto two of the colors.  For instance, one can form a projected
bipartite network of red and green vertices, in which a red and a green
vertex are connected by an ordinary edge if they were connected by a
hyperedge in the original network.  Thus one can create a network of, for
example, photos and the tags applied to them, while dropping information
about which users applied which tags.  And again one can also construct the
equivalent projections onto red/blue and blue/green vertex combinations.
Alternatively, one can construct a red/green network by connecting any pair
of vertices---of different colors or not---if they share a common blue
neighbor.  Thus a tag would be connected to a photo if any user applied that
tag to that photo, but tags would also be connected to other tags that were
used by the same user.

Many other standard concepts in the theory of networks can be generalized
to tripartite graphs, including clustering coefficients, correlations
between the degrees of adjacent vertices (including three-point
correlations), community structure and modularity, motif counts, and more.
The concepts introduced above, however, will be sufficient for our purposes
in this paper.

\section{Random tripartite graphs}
\label{sec:randgraph}
In theoretical studies of networks, random graph models have received
particular emphasis because they capture many of the essential properties
of networked systems in the real world while simultaneously being amenable
to analytic treatment.  A variety of random graph models have been studied,
from models of simple undirected or directed graphs to more complicated
examples with correlations, communities, or bipartite
structure~\cite{ER60,MR95,Bollobas01,NSW01,Newman02f}.  In this section we
develop the theory of random tripartite hypergraphs with given degree
distributions, which turn out to model many of the properties of real
tripartite graphs quite effectively.

\subsection{The model}
\label{sec:model}
Consider a model hypergraph with $n_r$~red vertices, $n_g$~green vertices,
and $n_b$~blue vertices.  Each vertex is assigned a degree, corresponding
to the number of hyperedges it will have.  These degrees can be visualized
as ``stubs'' of hyperedges emerging from each vertex in the appropriate
numbers.  The degrees must satisfy Eq.~\eqref{eq:constraint}, so that the
total number of stubs emerging from vertices of each color is the same and
equal to the total desired number of hyperedges~$m$.

A total of~$m$ three-way hyperedges are now created by choosing trios of
stubs uniformly at random, one each from a red, green, and blue vertex, and
connecting them to form hyperedges.  This model is the equivalent for our
tripartite graph of the so-called ``configuration model'' for unipartite
graphs~\cite{MR95} and the random bipartite graph model of~\cite{NSW01} for
bipartite graphs.

Given the definition of the model, we can, for example, calculate the
probability that a hyperedge exists between a given trio of vertices
$i,j,k$.  In the process of creating a single hyperedge, the probability
that we will choose a specific stub attached to red vertex~$i$ is $1/m$,
since there are a total of $m$ stubs attached to red vertices and we choose
uniformly among them.  If $i$ has degree~$k_i$ then the total probability
of choosing a stub from vertex~$i$ is $k_i/m$.  Similarly the probability
of choosing stubs from green and blue vertices $j$ and $k$ are $k_j/m$ and
$k_k/m$.  Given that there are $m$ hyperedges in total, the overall
probability of a hyperedge between $i$, $j$, and~$k$ is then
\begin{equation}
P_{ijk} = m\times{k_i\over m}\times{k_j\over m}\times{k_k\over m}
        = {k_ik_jk_k\over m^2}.
\label{eq:pijk}
\end{equation}
Via a similar argument, the probability that there is a hyperedge
connecting a particular red/green pair $i,j$ (or any other color
combination) is $k_i k_j/m$.  Note that in a sparse graph in which the
typical degrees remain constant as the size of the graph increases, both of
these probabilities vanish as $1/m$.  Among other things, this implies that
the chance of occurrence of small loops in the network vanishes in the
limit of large graph size.  In the language of graph theory, one says that
the network is \defn{locally tree-like}, a property that will be important
in the developments to follow.

Rather than specifying the degree of every vertex in the network, we can
alternatively specify just the degree distributions $p_r(k)$, $p_g(k)$,
and~$p_b(k)$ of the three vertex types (constrained to satisfy the sum
rules \eqref{eq:sumrule1} and~\eqref{eq:sumrule2}), then draw a specific
sequence of degrees from those distributions and connect the vertices as
before.  As a practical matter, if one wanted to generate actual example
networks on a computer, one would need to ensure that the degrees satisfied
Eq.~\eqref{eq:constraint}, which in general they will not on first being
drawn from the distributions.  A simple strategy for ensuring that they do
is first to draw a complete set of degrees and then repeatedly choose at
random a trio of vertices, one of each color, discard the current values of
their degrees, and redraw them from the appropriate distributions until the
constraint is satisfied.

The degree distributions represent the probability that a vertex of a given
color chosen at random from the entire network has a given degree.  If we
choose a \emph{hyperedge} at random, however, and follow it to the red,
green, or blue vertex at one of its corners, that vertex will not have
degree distributed according to $p_r(k)$, $p_g(k)$, or~$p_b(k)$, and the
reason is easy to see: vertices with many hyperedges are proportionately
more likely to be encountered when following edges.  A vertex of degree
ten, for instance, has ten times as many chances to be chosen in this way
than a similarly colored vertex of degree one.  (And a vertex of degree
zero will never be chosen at all.)  Thus the distribution of degrees of
vertices encountered is proportional to $kp_r(k)$ for red vertices, and
similarly for green and blue.  Requiring this distribution to sum to unity,
the correctly normalized distribution is $kp_r(k)/\sum_k kp_r(k) =
kp_r(k)/c_r$.

As in other random graph models, we are in fact usually interested not in
the degree of the vertex we encounter but in the number of hyperedges
attached to it other than the one we followed to reach it.  This so-called
\defn{excess degree}, which is 1 less than the total degree, has the same
distribution as above, but with the replacement $k\to k+1$, giving an
\defn{excess degree distribution} of
\begin{equation}
q_r(k) = {(k+1) p_r(k+1)\over c_r},
\label{eq:defsqr}
\end{equation}
and similarly for other vertex colors.

\subsection{Generating functions}
The fundamental tools we will use in calculating the properties of the
random tripartite graph are probability generating functions.  We begin by
defining generating functions for the degree distributions thus:
\begin{subequations}
\begin{align}
r_0(z) &= \sum_{k=0}^\infty p_r(k) z^k, \\
g_0(z) &= \sum_{k=0}^\infty p_g(k) z^k, \\
b_0(z) &= \sum_{k=0}^\infty p_b(k) z^k.
\end{align}
\label{eq:gen1}
\end{subequations}
Given these generating functions we can, for instance, easily calculate the
means of the distributions: $c_r=r_0'(1)$ and so forth.  Higher moments are
also straightforward.

We also define corresponding generating functions for the excess degree
distributions:
\begin{subequations}
\begin{equation}
r_1(z) = \sum_{k=0}^\infty q_r(k) z^k
        = {1\over c_r} \sum_{k=0}^\infty (k+1) p_r(k+1) z^k
        = {r_0'(z)\over r_0'(1)},
\end{equation}
and
\begin{align}
g_1(z) &= \sum_{k=0}^\infty q_g(k) z^k
        = {g_0'(z)\over g_0'(1)}, \\
b_1(z) &= \sum_{k=0}^\infty q_b(k) z^k
        = {b_0'(z)\over b_0'(1)}.
\end{align}
\label{eq:gen2}
\end{subequations}

\subsection{Projections}
\label{sec:projns}
As a first example, we use our generating functions to calculate the degree
distribution for the projection of a tripartite random graph onto one of
its vertex types, as described in Section~\ref{sec:tripartite}.  Consider
first the projection onto (say) red vertices in which two red vertices are
joined by an edge if they share a green neighbor.  (The blue vertices are
ignored in this projection.)

Suppose a given red vertex~A has $s$ green neighbors and each of those
green neighbors has $t$ red neighbors other than vertex~A.  Given that $s$
is distributed according to~$p_r(s)$ and $t$ is distributed according
to~$q_g(t)$, the probability~$\rho_g(k)$ that A has exactly $k$ neighbors
in the projected network is
\begin{equation}
  \rho_g(k) = \sum_{s=0}^\infty p_r(s)
  \sum_{t_1=0}^\infty q_g(t_1) \ldots \sum_{t_s=0}^\infty q_g(t_s)
  \,\delta\biggl( k, \sum_{n=1}^s t_n \biggr),
\end{equation}
where $\delta(i,j)$ is the Kronecker delta.  Multiplying both sides by
$z^k$ and summing over~$k$, the generating function for this probability
distribution is,
\begin{align}
R_g(z) &= \sum_{k=0}^\infty z^k \sum_{s=0}^\infty p_r(s) \nonumber\\
       &  \qquad {}\times
          \sum_{t_1=0}^\infty q_g(t_1) \ldots \sum_{t_s=0}^\infty q_g(t_s)
          \,\delta\biggl( k, \sum_{n=1}^s t_n \biggr) \nonumber\\
       &= \sum_{s=0}^\infty p_r(s)
          \sum_{t_1=0}^\infty q_g(t_1) \ldots \sum_{t_s=0}^\infty q_g(t_s)
          z^{\sum_n t_n} \nonumber\\
       &= \sum_{s=0}^\infty p_r(s)
          \sum_{t_1=0}^\infty q_g(t_1) z^{t_1} \ldots
          \sum_{t_s=0}^\infty q_g(t_s) z^{t_s} \nonumber\\
       &= \sum_{s=0}^\infty p_r(s)
          \Biggl[ \sum_{t=0}^\infty q_g(t) z^t \Biggr]^s
        = \sum_{s=0}^\infty p_r(s) \bigl[ g_1(z) \bigr]^s \nonumber\\
       &= r_0(g_1(z)).
\label{eq:defsrg}
\end{align}

We can also calculate the generating function for the projection in which
two red vertices are connected by an edge if they share either a green or a
blue neighbor.  The probability for a vertex to have~$k$ neighbors in this
network is
\begin{align}
\rho_{gb}(k) &= \sum_{s=0}^\infty p_r(s)
          \sum_{t_1=0}^\infty q_g(t_1) \ldots \sum_{t_s=0}^\infty q_g(t_s)
          \nonumber\\
       &  \quad{}\times
          \sum_{u_1=0}^\infty q_b(u_1) \ldots \sum_{u_s=0}^\infty q_b(u_s)
          \,\delta\biggl( k, \sum_{n=1}^s (t_n+u_n) \biggr),
\end{align}
and the corresponding generating function is
\begin{align}
R_{gb}(z) &= \sum_{k=0}^\infty z^k \sum_{s=0}^\infty p_r(s)
        \sum_{t_1=0}^\infty q_g(t_1) \ldots \sum_{t_s=0}^\infty q_g(t_s)
        \nonumber\\
     &  \quad{}\times
        \sum_{u_1=0}^\infty q_b(u_1) \ldots \sum_{u_s=0}^\infty q_b(u_s)
        \,\delta\biggl( k, \sum_{n=1}^s (t_n+u_n) \biggr) \nonumber\\
     &= \sum_{s=0}^\infty p_r(s)
        \Biggl[ \sum_{t=0}^\infty q_g(t) z^t \Biggr]^s
        \Biggl[ \sum_{u=0}^\infty q_b(u) z^u \Biggr]^s \nonumber\\
     &= r_0( g_1(z) b_1(z) ).
\label{eq:defsrgb}
\end{align}

We can use this result to calculate, for instance, the average degree in
the projected network, which is given by
\begin{equation}
R'_{gb}(1) = r'_0(1) \bigl[ b'_1(1) + g'_1(1) \bigr].
\end{equation}
We will also use it in Section~\ref{sec:apps} to compare predictions of the
random graph model with real-world networks.

\subsection{Formation and size of the giant component}
\label{sec:gcomp}
In this section we examine the component structure of our model network,
focusing on the giant component.  As with all networks, if our tripartite
network is sufficiently sparse---if it has very few edges for the given
number of vertices---then vertices will be connected together only in small
groups or \defn{small components}.  If, however, the number of edges is
sufficiently high, then a fraction of the vertices will join together into
a single large group, the \defn{giant component}, with the remainder in
small components.  There is a phase transition with increasing density at
which the giant component forms that is closely analogous to the phase
transition in classical percolation.

There is more than one possible definition of a component in our tripartite
network, but the simplest approach is to define it as a set of vertices of
any colors that are connected via hyperedges such that every vertex in the
set is reachable from every other by some path through the network.  Thus
the collection of vertices depicted in the top panel of
Fig.~\ref{fig:project} constitutes a component in this sense.

When viewed in the context of folksonomies, components, and particularly
the giant component, play an important practical role.  In a folksonomy
such as that of Flickr, the photography web site, users can ``surf''
between photographs by traversing the hypergraph.  A user can, for example,
click on the tag associated with a photo and see a list of other photos
with the same tag.  Similarly a user can click on the name of another user
and see a list of photos that user has tagged.  The existence, or not, of a
giant component in the network dictates whether this type of surfing is
actually useful or not.  If there is no giant component, then surfing users
will find themselves restricted to the small set of photos, tags, and users
in the component in which they start their surfing.  But if there is a
giant component then users will be able to surf to a significant fraction
of all photos on the entire web site just by clicking on tags or users that
seem interesting.  The same considerations affect automated surfing by
computerized ``crawlers'' that crawl web sites either to perform directed
searches (so-called ``spiders'') or to create indexes for later search.  If
there is no giant component in the folksonomy, then it cannot be crawled in
a useful way.

\begin{figure}[t]
\includegraphics[width=3.5cm]{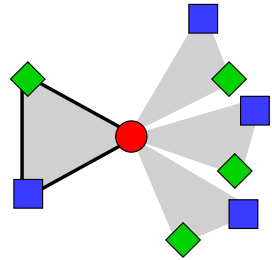}
\caption{If a hyperedge (outlined in bold) is not to belong to the giant
  component, then it must be that none of the hyperedges reachable via, for
  instance, its red vertex are themselves members of the giant component.}
\label{fig:gc}
\end{figure}

We can calculate properties of the giant component in our tripartite random
graph by methods similar to those used for ordinary random
graphs~\cite{NSW01}.  Consider a randomly chosen hyperedge in the full
hypergraph, as depicted in Fig.~\ref{fig:gc}, and let us calculate the
probability that this hyperedge is \emph{not} a part of the giant
component.  We define $u_r$ to be the probability that the hyperedge is not
connected to the giant component via its red vertex, and similarly for
$u_g$ and~$u_b$, so that the total probability of not belonging to the
giant component is $u_ru_gu_b$.

Suppose that the excess degree of the red vertex---the number of other
hyperedges attached to it---is~$k$.  (In the example shown in
Fig.~\ref{fig:gc} we have $k=3$.)  In order that the hyperedge be not
connected to the giant component via the red vertex it must be that none of
these other hyperedges are connected to the giant component either.  Any
one hyperedge satisfies this criterion with probability $u_gu_b$---the
probability that neither of its other corners lead to the giant
component---and all $k$ of them together do so with probability
$(u_gu_b)^k$.

The excess degree is distributed according to the distribution~$q_r(k)$
defined in Eq.~\eqref{eq:defsqr}.  Averaging over this distribution, we
then derive an expression for~$u_r$ thus:
\begin{equation}
u_r = \sum_{k=0}^\infty q_r(k) (u_gu_b)^k = r_1(u_gu_b).
\label{eq:solveur}
\end{equation}
Similarly we can show that
\begin{equation}
u_g = g_1(u_bu_r),\qquad u_b = b_1(u_ru_g).
\label{eq:solveugb}
\end{equation}
The simultaneous solution of these three equations for $u_r$, $u_g$,
and~$u_b$ then allows us to calculate the probability $1-u_ru_gu_b$ that a
randomly chosen hyperedge is in the giant component.  Alternatively, the
probability that a randomly chosen red vertex is not in the giant component
is the probability that none of its $k$ hyperedges lead to the giant
component, which is $\sum_k p_r(k) (u_gu_b)^k = r_0(u_gu_b)$, so the that a
red vertex \emph{is} in the giant component with probability
\begin{equation}
S_r = 1 - r_0(u_gu_b),
\end{equation}
and we can write similar equations for $S_g$ and~$S_b$.  $S_r$~can also be
thought of as the fraction of red vertices in the giant component, and
hence is a measure of the size of that component.  The absolute number of
red vertices in the giant component is $n_r S_r$ and the number of vertices
of all colors is $n_r S_r + n_g S_g + n_b S_b$.

As in other random graph models, it is in most cases not possible to solve
Eqs.~\eqref{eq:solveur} and~\eqref{eq:solveugb} for $u_r$, $u_g$, and~$u_b$
in closed form, but a numerical solution can be found easily by iteration
starting from suitable initial values.

We can also derive a condition for the existence of a giant component in
the network.  A giant component exists if and only if $u_r$, $u_g$, and
$u_b$ are all less than~1.  (They must all be less than~1 because an
extensive giant component of vertices of any one color automatically
implies an extensive component of the other two colors, since, with only
mild conditions on the degree distribution, the first color must be
connected into a giant component by an extensive number of hyperedges, and
each hyperedge is attached to one vertex of each color.)

Consider values of the variables that are only slightly different from~1
thus:
\begin{equation}
u_r = 1 - \epsilon_r,\qquad
u_g = 1 - \epsilon_g,\qquad
u_b = 1 - \epsilon_b,
\end{equation}
where $\epsilon_r$, $\epsilon_g$, and $\epsilon_b$ are small.  Then, from
Eq.~\eqref{eq:solveur},
\begin{align}
\epsilon_r &= 1 - u_r = 1 - r_1(u_gu_b)
     = 1 - r_1(1-\epsilon_g-\epsilon_b+\epsilon_g\epsilon_b) \nonumber\\
    &= (\epsilon_g+\epsilon_b) r_1'(1) + \Ord(\epsilon^2),
\end{align}
where we have performed a Taylor expansion of $r_1$ and made use of
$r_1(1)=1$ (which is necessarily true if $q_r(k)$ is a properly normalized
distribution).  We can derive similar equations for $\epsilon_g$ and
$\epsilon_b$ and combine all three into the single vector equation
\begin{equation}
\setlength{\arraycolsep}{4pt}
\begin{pmatrix}
  \epsilon_r \\ \epsilon_g \\ \epsilon_b
\end{pmatrix}
= \begin{pmatrix}
  0 & r & r \\
  g & 0 & g \\
  b & b & 0
\end{pmatrix}
\begin{pmatrix}\epsilon_r \\ \epsilon_g \\ \epsilon_b
\end{pmatrix},
\label{eq:perturb1}
\end{equation}
where we have introduced the shorthand $r=r_1'(1)$, $g=g_1'(1)$, and
$b=b_1'(1)$.

If $u_r$, $u_g$, and $u_b$ are to be less than 1, meaning the corresponding $\epsilon$'s must all be non-zero,  then this equation implies
the determinant condition
\begin{equation}
\begin{vmatrix}
  -1 & r & r \\
  g & -1 & g \\
  b & b & -1
\end{vmatrix} = 0,
\end{equation}
or
\begin{equation}
2rgb + rg + gb + br = 1.
\end{equation}

This condition defines the point at which the phase transition takes place.
Equivalently, $2rgb + rg + gb + br$ crosses~1 at the transition.  In fact
it is greater than~1 when there is a giant component and less~1 when there
is none (rather than the other way around) as can be shown by exhibiting
any example where this is the case.  A suitable example is provided by a
network in which all vertices have degree one, which clearly has no giant
component.  This choice makes $r=g=b=0$ and the result follows.

Thus our condition for the existence of a giant component is,
\begin{equation}
2rgb + rg + gb + br > 1.
\label{eq:rgbcond}
\end{equation}
This is the equivalent of the well known condition of Molloy and Reed for
the existence of a giant component in a unipartite random
graph~\cite{MR95}.

An alternative form for this condition can be derived by making use of
Eqs.~\eqref{eq:defsqr} and~\eqref{eq:gen2} to write
\begin{align}
r &= r_1'(1) = \sum_{k=0}^\infty k q_r(k)
         = {1\over c_r} \sum_{k=0}^\infty k(k+1) p_r(k+1) \nonumber\\
        &= {1\over c_r} \sum_{k=0}^\infty k(k-1) p_r(k)
         = {\av{k^2}_r\over\av{k}_r} - 1,
\label{eq:valr}
\end{align}
and similarly for $g$ and~$b$.  Here $\av{\ldots}_r$ indicates an average
over the degree distribution of the red vertices and $c_r=\av{k}_r$.

Substituting these expressions into~\eqref{eq:rgbcond}, we find, after some
algebra, that
\begin{equation}
{\av{k}_r\over\av{k^2}_r} + {\av{k}_g\over\av{k^2}_g} + 
{\av{k}_b\over\av{k^2}_b} < 2.
\label{eq:rgbcriterion}
\end{equation}
This form is particularly pleasing, since it has the same general shape as
the criterion of Molloy and Reed for the unipartite case, which can be
written as $\av{k}/\av{k^2} < \frac12$.

\subsection{Other types of components}
The definition of a component used in the previous section is not the only
one possible for our tripartite graph.  In some folksonomies one cannot
surf over connections formed by both users and tags.  In some cases, for
instance, one is barred from seeing which resources a particular user has
tagged for privacy reasons, meaning one can surf between resources with the
same tag, but not with the same user.  In this case we are surfing on the
network formed by two colors of vertices only, say red and green.

We can approach this situation using the same techniques as in the previous
section.  We define probabilities $u_r$ and $u_g$ as before and find that
they satisfy the equations,
\begin{equation}
u_r = r_1(u_g),\qquad u_g = g_1(u_r).
\end{equation}
Linearizing around the point $u_r=u_g=1$ we then find that the transition
at which the giant component appears takes place when
\begin{equation}
\begin{vmatrix} -1 & r \\ g & -1 \end{vmatrix} = 0,
\end{equation}
or equivalently $rg=1$, with $r$ and $g$ defined as before.  By considering
appropriate special cases, one can then show that the giant component
exists if and only if $rg>1$.  Substituting from Eq.~\eqref{eq:valr}, we
can also write this condition in the form,
\begin{equation}
{\av{k}_r\over\av{k^2}_r} + {\av{k}_g\over\av{k^2}_g} < 1.
\end{equation}
Note that this expression is not symmetric with respect to permutations of
the three color indices, as Eq.~\eqref{eq:rgbcriterion} was.  This means
that in general giant components for different color pairs will appear at
different transitions, and it is possible to have a giant component for one
pair without having a giant component for another.  Thus for instance in
our Flickr example one might be able to surf the network of photos and
tags, but not the network of photos and users.  (Actually, one can surf
both just fine in the real Flickr network.)

\subsection{Percolation}
\label{sec:perc}
One can also consider percolation processes on tripartite networks.  If
some vertices are removed from the network then the remaining network may
or may not percolate, i.e.,~possess a giant component.  For example, on the
Flickr web site users can designate photos as publicly viewable or not, and
those that are not are, for all intents and purposes, removed from the
network.  One cannot use them, for instance, for surfing across the
network.  There are many ways in which vertices might be removed, but as a
simple example let us assume that vertices of only one kind are removed and
make the standard percolation assumption that they are removed uniformly at
random.  (More complicated percolation schemes are certainly possible, with
more than one type of vertex removed, different probabilities of removal
for different types, or nonuniform removal, and all of these schemes can be
studied by methods similar to those outlined here.)

Suppose a fraction $\phi$ of the red vertices in our network are present
(or functional) and $1-\phi$ are removed (or nonfunctional).  In the
language of percolation theory, a fraction $\phi$ of the vertices are
occupied.  Then define $u_r$ as before to be the probability that the red
vertex attached to a random hyperedge does not belong to the giant
component, or the \defn{giant cluster} as it is more commonly called in the
percolation context.  There are two different ways in which this can
happen.  If the vertex itself has been removed, then it does not belong to
the giant cluster.  Alternatively, it may be present but, as before, none
of its neighbors, either blue or green, are in the giant cluster.  This
allows us to write down an expression for $u_r$ thus:
\begin{equation}
u_r = 1 - \phi + \phi r_1(u_gu_b).
\label{eq:perc1}
\end{equation}
The corresponding expressions for $u_g$ and $u_b$ are the same as in our
previous calculation, $u_g = g_1(u_bu_r)$, $u_b = b_1(u_ru_g)$, and the
fractions of red, green, and blue vertices in the giant percolation cluster
are
\begin{subequations}
\begin{align}
S_r &= \phi[ 1 - r_0(u_gu_b) ],\\
S_g &= 1 - g_0(u_bu_r),\\
S_b &= 1 - b_0(u_ru_g).
\end{align}
\label{eq:perc3}
\end{subequations}

We can also calculate an expression for the value of $\phi$ at which the
percolation transition happens.  As before we perturb around the point
$u_r=u_g=u_b=1$ that corresponds to no giant cluster and the equivalent of
Eq.~\eqref{eq:perturb1} is
\begin{equation}
\setlength{\arraycolsep}{4pt}
\begin{pmatrix}
  \epsilon_r \\ \epsilon_g \\ \epsilon_b
\end{pmatrix}
= \begin{pmatrix}
  0 & \phi r & \phi r \\
  g & 0 & g \\
  b & b & 0
\end{pmatrix}
\begin{pmatrix}\epsilon_r \\ \epsilon_g \\ \epsilon_b
\end{pmatrix},
\end{equation}
with $r$, $g$, and $b$ defined as before.  This implies that the transition
happens at $\phi=\phi_c$ where $\phi_c$ is the solution of $2\phi rgb +
\phi rg + gb + \phi br = 1$.  That is,
\begin{equation}
\phi_c = {1 - gb\over r(2gb+g+b)}.
\end{equation}
Making use of Eq.~\eqref{eq:valr} and the corresponding expressions for $g$
and $b$ we then find that
\begin{equation}
\phi_c = \biggl( {\av{k^2}_r\over\av{k}_r} - 1 \biggr)^{-1}
         \biggl[
         \biggl( 2 - {\av{k}_g\over\av{k^2}_g} - {\av{k}_b\over\av{k^2}_b}
         \biggr)^{-1} - 1 \biggr].
\end{equation}

\subsection{Simulations}
\label{sec:example}
Before looking at real-world tripartite networks, we first compare our
calculations with simulation results for computer-generated random graphs.

Consider a tripartite random graph with Poisson degree distributions thus:
\begin{equation}
p_r(k) = \e^{-c_r} {c_r^k\over k!},\quad
p_g(k) = \e^{-c_g} {c_g^k\over k!},\quad
p_b(k) = \e^{-c_b} {c_b^k\over k!},
\end{equation}
where the average degrees $c_r$, $c_g$, and $c_b$ satisfy
Eq.~\eqref{eq:constraint}.  The corresponding generating functions are
\begin{align}
r_0(z) &= r_1(z) = \e^{-c_r} \sum_{k=0}^\infty \frac{c_r^k}{k!} z^k
        = \e^{c_r(z-1)}, \nonumber\\
g_0(z) &= g_1(z) = \e^{-c_g} \sum_{k=0}^\infty \frac{c_g^k}{k!} z^k
        = \e^{c_g(z-1)}, \nonumber\\
b_0(z) &= b_1(z) = \e^{-c_b} \sum_{k=0}^\infty \frac{c_b^k}{k!} z^k
        = \e^{c_b(z-1)}.
\label{eq:glist1}
\end{align}
We can use these to calculate, for instance, the degree distribution of the
projection of the network onto the red vertices in which two vertices are
connected if they share either a green or a blue neighbor.  The generating
function for this distribution is given by Eq.~\eqref{eq:defsrgb} to be
\begin{equation}
R_{gb} = r_0(g_1(z)b_1(z)) = \e^{c_r(\e^{(c_g+c_b)(z-1)}-1)}.
\end{equation}
Expanding in powers of~$z$, we then find that the
probability~$\rho_{gb}(k)$ of a red vertex having exactly~$k$ neighbors in
the projected network is
\begin{align}
\rho_{gb}(k) &= {(c_g+c_b)^k\over k!} \e^{c_r(\e^{-(c_g+c_b)}-1)} \nonumber\\
  & \qquad{}\times
    \sum_{m=1}^k \stirling{k}{m} \bigl[ c_r\e^{-(c_g+c_b)} \bigr]^m,
\label{eq:redpi}
\end{align}
where $\bigl\lbrace{k\atop m}\bigr\rbrace$ is a Stirling number of the
second kind, i.e.,~the number of ways of dividing $k$ objects into $m$
nonempty sets~\cite{AS65}.

\begin{figure}[t]
\includegraphics[width=\columnwidth]{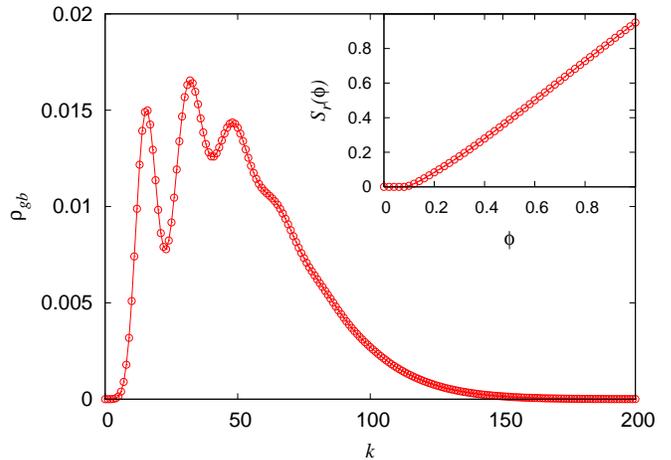}
\caption{The degree distribution for the projection of our Poisson
  hypergraph onto its red vertices alone, in which two red vertices are
  joined by an edge if they have either a green or a blue neighbor in
  common on the original tripartite network.  The solid line is the exact
  solution, Eq.~\eqref{eq:redpi}, and the points are the results of
  numerical simulations averaged over a hundred realizations of the
  network.  The error bars are smaller than the size of the points in all
  cases.  Inset: The fraction of red vertices belonging to the giant
  percolation cluster for site percolation on the tripartite network, as a
  function of occupation probability~$\phi$.  The solid line is the exact
  solution and the points are the results of numerical simulations.}
\label{fig:simulations}
\end{figure}

The main panel of Fig.~\ref{fig:simulations} shows the form of this
distribution for the case $c_r=3$, $c_g=10$, $c_b=6$.  In the same plot we
show the results of simulations in which random tripartite graphs with the
same degree distributions and $n_r=100\,000$, $n_g=30\,000$, and
$n_b=50\,000$ were generated and then explicity projected onto the red
vertices and the resulting degree distribution measured directly.  As the
figure shows, the agreement between the two is excellent.

The inset of Fig.~\ref{fig:simulations} shows the size of the giant cluster
for percolation on the red vertices of the same network as a function of
the occupation probability~$\phi$, calculated both by numerical solution of
Eqs.~\eqref{eq:perc1}--\eqref{eq:perc3} and by direct measurement on
simulated networks.  Again the agreement is excellent.

\section{Comparison with real-world data}
\label{sec:apps}
In this section we compare the predictions of our tripartite random graph
model against data for the folksonomy of the Flickr photo-sharing web site.
As we show, the theory and empirical observations agree well in some
respects, but less well in others.  In many ways the discrepancies are at
least as interesting as the cases of agreement, since they indicate
situations in which the structure of the observed network cannot be
explained by a simple random model that ignores social and other effects.
When data and model disagree it is a sign that these effects are important
in determining the network structure.  Thus, as with other random graph
models, one of the most significant roles our model can play may be as a
null model that allows the experimenter to determine when a network is
doing something nontrivial.

\begin{figure}[t]
\includegraphics[width=9cm]{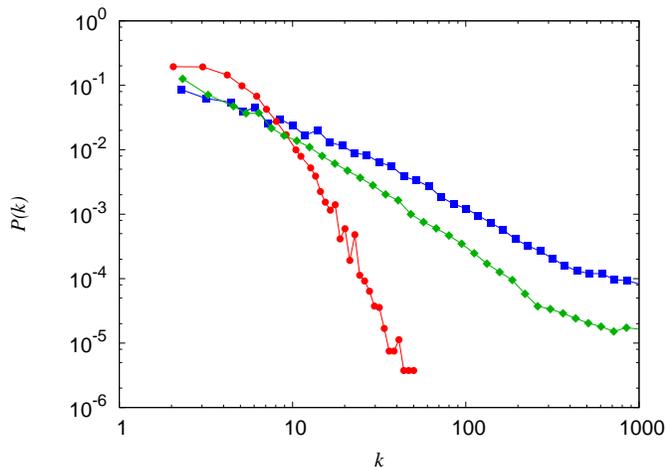}
\caption{The three degree distributions of the tripartite Flickr folksonomy
  network for photos (red), tags (green), and users (blue).}
\label{fig:3deg}
\end{figure}

Our example data set represents the folksonomy network of $266\,198$ photos
added to the Flickr web site by its users during 2007, along with the tags
applied to those photos and the users who applied them.  The first step in
analyzing the data is to measure the three degree distributions for the
three types of vertices.  The degree distributions are shown in
Fig.~\ref{fig:3deg}.  As is common in most social networks, they are highly
right-skewed, meaning there are many vertices of low degree and a small
number of very high degree, although the distributions do not follow
power-law forms as the distributions in some networks do.  Using these
distributions, we can, following Eqs.~\eqref{eq:gen1} and \eqref{eq:gen2},
construct the corresponding generating functions, which are simple
polynomials (albeit of high order) that can be easily evaluated
numerically.

We can use our generating functions to calculate, for example, the
generating functions~$R_{gb}(z)$ and so forth for the degree distributions of
the projections of the network onto one vertex type, using
Eqs.~\eqref{eq:defsrg} and~\eqref{eq:defsrgb} and their equivalents for
other vertex types.  Again these functions can be rapidly evaluated for any
argument~$z$ numerically.  The degree distributions themselves are then
given by derivatives of the generating functions thus:
\begin{equation}
p_k = \frac{1}{k!}\frac{\d^k R_{gb}}{\d z^k}\biggr|_{z=0}.
\label{eq:coeffs}
\end{equation}
Direct numerical evaluation of derivatives is plagued by problems with
noise and should be avoided, but one can get good results~\cite{MN00b} by
instead employing Cauchy's integral formula for the $k$th derivative of a
function:
\begin{equation}
\frac{\d^k\!f}{\d z^k} \biggr|_{z=z_0}
  = \frac{k!}{2\pi\ii} \oint \frac{f(z)}{(z-z_0)^{k+1}} \>\d z,
\label{eq:cauchy}
\end{equation}
where the integral is around a contour enclosing the point $z_0$ but
excluding any poles of $f(z)$.  Applying this formula to~\eqref{eq:coeffs}
we get
\begin{equation}
p_k = \frac{1}{2\pi\ii} \oint \frac{R_{gb}(z)}{z^{k+1}} \>\d z.
\label{eq:numpk}
\end{equation}
We then calculate the degree distribution by performing the contour
integral numerically around a suitable contour (the unit circle $|z|=1$
works well).  One can without difficulty calculate to good precision the
first thousand or so coefficients of the generating function in this
fashion.

\begin{figure}[t]
\includegraphics[width=9cm]{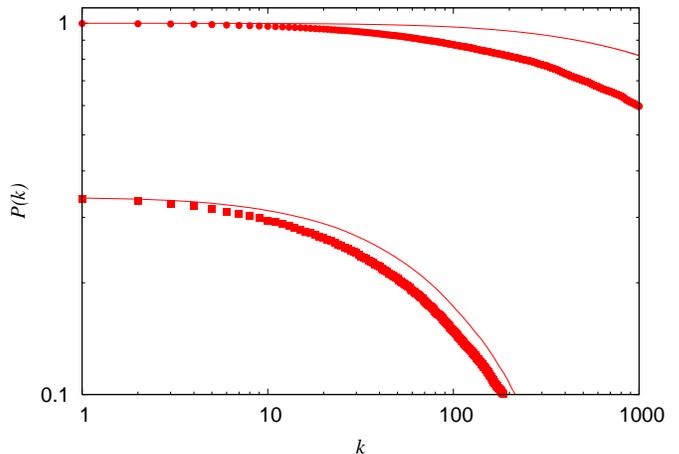}
\caption{Circles show the cumulative distribution function for the degree
  distribution of the projection of the Flickr network onto its photograph
  vertices, while the upper solid line shows the predictions of the random
  graph model for the same quantity.  Squares show the same function after
  pruning of the data to remove multiple tagging as described in the text
  and the lower solid curve shows the corresponding model prediction,
  recalculated from the new degree distributions after pruning.}
\label{fig:usercdf}
\end{figure}

We have performed this calculation using the degree distributions of the
Flickr network and projecting onto the resources, i.e.,~the photos.
Figure~\ref{fig:usercdf} shows a comparison of the results with the degree
distribution for the actual projected network.  The upper solid line in the
figure represents the theoretical result, while the circles represent the
measurements.  Although the two curves have the same general shape, it's
clear from the figure that the agreement between them is only moderately
good in this case.  Upon closer inspection, however, it turns out that
there is a relatively simple reason for this.

As discussed in Section~\ref{sec:model}, our random graph model assumes a
locally-tree like structure for the tripartite network, a structure with no
short loops.  The Flickr network, on the other hand, turns out to have many
short loops, which is why empirical measurements and model do not agree in
Fig.~\ref{fig:usercdf}.  As we now show, however, the loops in the Flickr
network are primarily of a trivial kind that can easily be allowed for in
the calculations.

Typically, photos are not added to the Flickr network individually, but in
sets.  The most common practice is for a user to upload a set of photos on
a particular subject---say, pictures of a Ferrari motor car---and then
label all of the photos in the set with the same set of
tags---\textit{Ferrari}, \textit{automobile}, \textit{sports car}, and so
forth.  This creates short loops between photos in the set of the form
$P_1\to T_1\to P_2\to T_2\to P_1$, where the $P$s are the photos and the
$T$s are tags.  These loops will have an adverse affect on the calculation
of the number of neighbors a photo has in the projected network, since in
many cases two projected edges from a photo will lead to the same
neighboring photo, rather than to different neighbors, and hence give a
lower degree in the projected network than our naive random graph
calculation.

To test the effect of these ``trivial'' loops in the network structure, we
have pruned the data set to remove instances of multiple tagging.  In the
pruned data set the application by a user of many tags to the same photo is
represented by just a single hyperedge, rather than many.  In this
representation, hyperedges represent the act of tagging a photo, rather
than a specific tag, and only one hyperedge is included between a user and
a photo no matter how many tags the user applies.  Similarly we also
represent the tagging of many photos with the same tag by a single
hyperedge, so that hyperedges represent the act of tagging an entire photo
set, rather than just a single photo.  This should remove most instances of
trivial loops in the projected network of the type described above.

Now we calculate again the projection of the hypergraph onto the set of
photos.  We also recalculate the theoretical predictions to reflect the
changed degree distributions of the hypergraph following pruning.  The
results are shown in Fig.~\ref{fig:usercdf} (squares and lower solid curve)
and, as the figure shows, the agreement is now quite good between theory
and observation.  This suggests that the earlier disagreement between the
two is indeed primarily a result of the presence of the loops in the
hypergraph introduced by the practice of multiple tagging.

\begin{figure}[t]
\includegraphics[width=9cm]{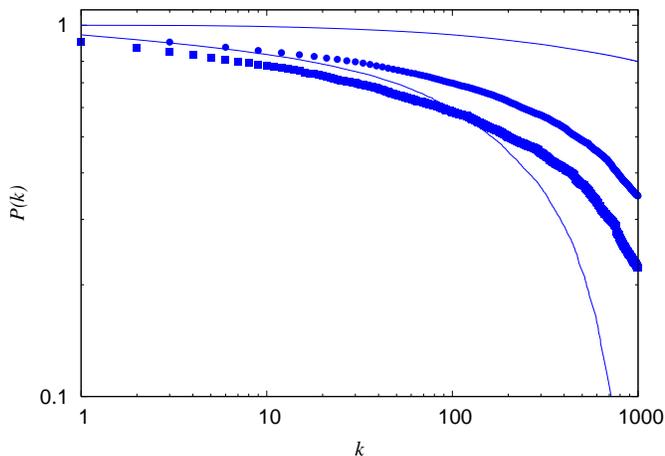}
\caption{Cumulative distribution functions for the degree distributions of
  the projection of the Flickr network onto its user vertices, both before
  and after pruning of the data.  The points represent the observations,
  unpruned (circles) and pruned (squares), while the solid lines represent
  the predictions of the model.}
\label{fig:photocdf}
\end{figure}

We can perform similar calculations for projections onto other types of
vertices.  In Fig.~\ref{fig:photocdf} we show degree distributions, before
and after pruning of the data set, for the projection onto users.
Agreement between theory and observation for the unpruned data is again
quite poor in this case but significantly better for the pruned data.

These calculations provide, in many ways, a good example of the utility of
random graph models.  When compared with the raw data from the Flickr
network, our random graph model agrees qualitatively, but not
quantitatively, indicating that there are effects present in the network
that are not accounted for by simple random hyperedges.  On the other hand,
once one prunes the data to remove multiple tagging, the agreement becomes
much better, suggesting that multiple tagging is the primary nonrandom
behavior taking place in the network and that in other respects the network
is in fact quite close to being a random graph.  Thus the model allows us
not only to say when the network deviates from the random assumption, but
also the particular nature of the deviation.

\section{Conclusions}
\label{sec:concs}
Motivated by the emergence of new types of social networks, such as
folksonomies, we have in this paper proposed and studied a model of random
tripartite hypergraphs.  We have defined basic network measures, such as
degree distributions and projections onto individual vertex types, and
calculated a variety of statistical properties of the model in the limit of
large network size.  Among other things we have calculated the explicit
degree distributions for projected networks, conditions for the emergence
of a giant component, the size of the giant component when there is one,
and the location of the percolation threshold for site percolation on the
network.  In principle, the techniques introduced could be extended to
hypergraphs with more vertex types or additional types of edges, although
we have not pursued any such extensions here.

We have compared our results against measurements of computer-generated
random hypergraphs and a real-world tripartite network, the folksonomy of
the on-line photo-sharing web site Flickr. In the latter case, we have
focused on the degree distributions of projections of the hypergraph onto
one vertex type and find that in some instances the theory makes
predictions in moderately good agreement with the observations while in
others the agreement is poorer.  In all cases, however, we find that
agreement becomes significantly better when we remove instances of multiple
tagging from the network---instances in which a user applies many tags to
the same photo or the same tag to many photos---suggesting that the
disagreement is primarily a result of relatively trivial structures in the
network, rather than more subtle or large-scale social network effects.

\begin{acknowledgments}
 The authors thank the EU TAGORA project for providing the Flickr dataset. This work was funded
  in part by the National Science Foundation under grant number
  DMS--0804778 and by the James S. McDonnell Foundation.
\end{acknowledgments}

\bibliographystyle{apsrev}
\bibliography{journals,references,randhyper2}

\end{document}